\documentclass[10pt,conference]{IEEEtran}

\usepackage{cite}
\usepackage{amsmath,amssymb,amsfonts}
\usepackage{algorithmic}
\usepackage{graphicx}
\usepackage{textcomp}
\usepackage[usenames,dvipsnames]{xcolor}
\usepackage{url}
\usepackage[hidelinks,bookmarks=false]{hyperref}

\usepackage{listings}
\usepackage{fancyvrb}
\usepackage{booktabs}

\usepackage{booktabs} 
\usepackage{graphicx}
\usepackage{xcolor}
\usepackage{gensymb}
\usepackage{hyperref}
\usepackage{booktabs}
\usepackage{graphicx}

\usepackage{multirow}
\usepackage[absolute]{textpos}

\def\BibTeX{{\rm B\kern-.05em{\sc i\kern-.025em b}\kern-.08em
		T\kern-.1667em\lower.7ex\hbox{E}\kern-.125emX}}

\makeatother

\begin{document}

\title{Calculating Software’s Energy Use and Carbon Emissions:  A Survey of the State of Art, Challenges, and the Way Ahead}

	\author{\IEEEauthorblockN{Priyavanshi Pathania$^\dagger$, Nikhil Bamby$^\dagger$, Rohit Mehra$^\dagger$, Samarth Sikand$^\dagger$, Vibhu Saujanya Sharma$^\dagger$, \\Vikrant Kaulgud$^\dagger$, Sanjay Podder$^\ddagger$, Adam P. Burden*}
	\IEEEauthorblockA{\textit{$^\dagger$Accenture Labs, India \hspace{0.4em}
			$^\ddagger$Accenture, India \hspace{0.4em}
			*Accenture, USA}\\
		{\{priyavanshi.pathania, nikhil.bamby, rohit.a.mehra, s.sikand, vibhu.sharma, vikrant.kaulgud, sanjay.podder, adam.p.burden\}}\\@accenture.com}
}
\maketitle

\begin{textblock*}{12cm}(1.5cm,26.2cm) 
	Accepted for publication at GREENS 2025 (co-located with ICSE 2025)
\end{textblock*}

\begin{abstract}

The proliferation of software and AI comes with a hidden risk: its growing energy and carbon footprint. As concerns regarding environmental sustainability come to the forefront, understanding and optimizing how software impacts the environment becomes paramount. In this paper, we present a state-of-the-art review of methods and tools that enable the measurement of software and AI-related energy and/or carbon emissions. We introduce a taxonomy to categorize the existing work as Monitoring, Estimation, or Black-Box approaches. We delve deeper into the tools and compare them across different dimensions and granularity—for example, whether their measurement encompasses energy and carbon emissions and the components considered (like CPU, GPU, RAM, etc.). We present our observations on the practical use (component wise consolidation of approaches) as well as the challenges that we have identified across the current state-of-the-art. As we start an initiative to address these challenges, we emphasize active collaboration across the community in this important field.

\end{abstract}
\section{Introduction}\label{introduction}

In recent years, technology and software systems have proliferated, significantly impacting our everyday lives. However, these software systems/software running on underlying hardware infrastructure have a carbon footprint of their own. The Information and Communication Technology (ICT) sector, which constitutes computing devices, data centers, communication networks, etc., currently contributes to 2.1\% -3.9\% of global greenhouse gas (GHG) emissions \cite{freitag2021real}. The latest trends suggest that without sustainable steps, ICT's GHG emissions could surpass 14\% of global 2016 GHG emissions by 2040 \cite{BELKHIR2018448}. This accounts for half of the current relative contribution of the transportation sector \cite{BELKHIR2018448}. 

Moreover, there has been exponential growth in Artificial Intelligence (AI) and Generative AI (Gen AI) based models, their use cases, and subsequently their share of carbon emissions. Studies suggest that training a single AI model can emit as much carbon as five cars in their lifetimes \cite{Strubell_Ganesh_McCallum_2020}. The average carbon emissions generated by AI models have increased by a factor of 100 (two orders of magnitude) from 2012 to 2021 \cite{luccioni2023counting}. Hence, there has been a growing emphasis on reducing software emissions, also highlighted in the \textit{Climate Change 2022: Mitigation of Climate Change} report \cite{ipcc}.

To manage this growing carbon footprint, optimizations are required across the software systems, the underlying hardware, or both. Software-related optimizations require active participation of end stakeholders in developing green software strategies. In fact, the 2023 State of Green Software (SOGS) report released by the Green Software Foundation (GSF) has highlighted that approximately 30\% of software practitioners are interested and involved in sustainability work while another 60\% are interested in these green initiatives \cite{sogs}. To support these stakeholders/practitioners, software energy and carbon calculation tools are required. Without appropriate calculation systems, it is challenging to evaluate the impact of green optimization strategies. For instance, replacing Azure's Standard\_D32s\_v4 VM running on very low
CPU utilization with Standard\_D16s\_v4 VM,
can potentially lead to 48\% savings in carbon emissions \cite{10628316}. Without energy and carbon calculators, it is difficult to understand these savings.

Hence, in this paper, we present a study of existing tools and approaches in the domain of energy and carbon calculation for software. Although some survey-based research has been published that compares tools across different parameters, but they are (i) limited in the number of tools they compare \cite{morand2024mlca, jay2023experimental}, (ii) lack a proper categorization for existing energy calculation approaches \cite{martiny2023towards}, (iii) do not explicitly highlight the challenges in the current state-of-the-art, (iv) lack a component-wise (CPU, GPU, RAM, Storage, Network) compilation of energy calculation approaches \cite{bannour-etal-2021-evaluating}, and (v) focus only on quantitative analysis of the tools. 

To overcome this void, we present our survey of 21 tools that allow energy and carbon calculation at different granularity (system, VM, container, etc.). In Section II, we propose a high-level taxonomy of calculation approaches that exist in the literature, along with their drawbacks. This is followed by a brief exploration of some tools that are frequently referenced in the green software community in Section III. We then compare and contrast the existing tools across multiple, carefully selected attributes, resulting in a \textit{Tool Matrix} in Section IV A. The taxonomy and further comparison (\textit{Tool Matrix}) enable a practitioner, novice or advanced, to easily select an energy calculation tool for her specific use case.  We also present a component-wise compilation of the calculation approaches (Section IV B) and existing challenges in the state-of-the-art (Section V) to encourage the green software community towards these future research areas.

\section{ Proposed Taxonomy}\label{taxonomy}

\textit{The Methodology:} We initiated the study by gathering a diverse range of energy and carbon calculation tools from sources such as academic literature, Cloud Service Providers (CSPs), the Green Software Foundation (GSF) repository, and open exploration of the space. We focused on the existing tools to uncover the underlying calculation approaches used in the state of the art. After a thorough analysis of the approaches based on the mechanism they employ for energy and carbon calculation, we developed a taxonomy as shown in Figure 1. At a high level, we divide the calculation approaches into: Monitoring, Estimation, and Black-Box based approaches. Before diving deeper, we want to emphasize a few parameters frequently discussed in the context of energy and carbon emissions of software systems.

\textbf{(i) Power:} A software code running on a device needs certain electricity and hence it consumes power (Watts). \textbf{(ii) Energy:} If this power consumption is multiplied by the time for which the code runs, we obtain energy consumption in Watt-hours or Kilowatt-hours. \textbf{(iii) Emissions:} Refers to the carbon emissions or the total GHG emissions released during the full cycle of software systems. \textbf{(iv) Operational Emissions:} Refers to the carbon  emissions generated from the running of software. \textbf{(v) Embodied Emissions:} Refers to the amount of carbon emitted across the creation and disposal of a hardware device, that is used to run software. \textbf{(vi) Regional Carbon Intensity (RCI)/ Grid Intensity:} Energy is a fundamental property, while Emissions is a derived quantity (based on RCI values). RCI is a measure of how much $CO_2$ equivalent emissions are produced per kilowatt-hour ($gCO_2eq/kWh$) in a particular location. \textbf{(vii) Software Carbon Intensity (SCI):}
 Proposed by GSF, SCI is a method for scoring a software system based on its carbon emissions \cite{gsfsci}. It can be per user, per device, etc. (rate of carbon emissions).

\begin{figure}[t]
	\centering
	\includegraphics[width=1.0\linewidth,scale=0.4]{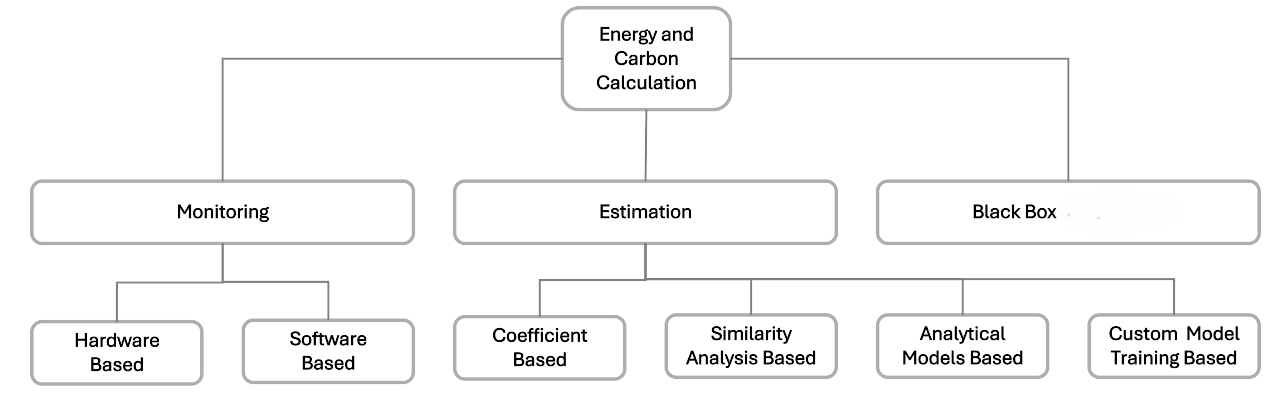}
	\caption{Proposed taxonomy of Energy and Carbon calculation approaches}
	\label{fig:taxonomy}
\end{figure}

\subsection{Monitoring Approaches}
Monitoring based approaches perform the actual measurement of energy using specific infrastructures or interfaces like sensors, monitors, and counters.

\subsubsection{Hardware Power Monitors}
These power monitors are characterized by physical power sensors that are deployed to measure the energy. The first category includes \textit{External Power Meters} which sit between the power supply and the machine (e.g., server) whose energy needs to be monitored. They provide the instantaneous energy measurement for the whole machine and are found to be highly accurate. By limiting the number of background processes being run on the machine, they can also measure the energy consumption of a particular application or process. Watts Up? Pro is one of the most popular power meters. Its wattage readings are found to be accurate within the range specified above 0.5 W \cite{hirst2013watts}. Other tools in this category include Kill-A-Watt, a wireless power meter PowerBlade, etc. \cite{killawatt}, \cite{debruin2015powerblade}. 
The next category is \textit{On-chip Power Monitors} which include devices like the Cotex-M0 based System on a chip (SoC) that can measure electricity signals and meter active and reactive energy \cite{8540878}.
\textbf{Limitations:}
Hardware monitoring tools are difficult to scale. With a growing number of servers, it is unfeasible to measure the corresponding energy through these power meters. Upgrading these tools to new versions is fairly difficult due to their hardware components \cite{noureddine2013review}.

\subsubsection{Software Power Meters}
They leverage hardware-centric libraries and energy counters accessed through model-specific registers (MSRs) to gather power-related metrics and calculate the energy at different granularity (machine to specific process). Some examples include Running Average Power Limit (RAPL) \cite{khan2018rapl}, Turbostat  \cite{turbostatlinux}, Intel Power Gadget \cite{prieto2022energy}, pyNVML \cite{pynvml}, etc. Software based power meters are widely used and have much larger scope than hardware based. Hence, in this paper, we explore more on software based power meters in Section III. \textbf{Limitations:} Most of these tools are not extensible to majority of cloud VMs (due to a lack of access to MSRs) and mainly support local machines and bare metal servers.

\subsection{Estimation Approaches}
These approaches take certain system parameters as input and aim to approximate the resulting energy using varied models, rules, patterns, etc. They are typically less intrusive than Monitoring approaches.

\subsubsection{Coefficient Based Estimation}
In this case, fixed coefficients, typically based on certain assumptions, are used to estimate energy. These static coefficients act as a predictor of how many watt-hours (Wh) are consumed per \textit{utilization} of a particular component. For instance, Cloud Jewels \cite{etsycloudjewels} estimates \textit{1.52 Wh/TBh for SSD storage} in a cloud environment. Static coefficients are relevant for almost all components like CPU, GPU, Memory, Storage, and Network. Another category of coefficient based approaches use Thermal Design Power (TDP) which refers to the maximum amount of heat generated by a component under a steady workload. It is typically represented in watts (W) and its actual goal is to guide system designers with a power target for accurate thermal solution selection \cite{inteltdp}. Existing research suggests that TDP can be used as an approximation of component power consumption at maximum use \cite{jay2023experimental}. Since Energy = Power * Time, CPU energy consumption is usually estimated as a product of TDP and the total execution time. Similarly, for GPUs, TDP can be assumed as its maximum Watt draw \cite{henderson2020towards}.
\textbf{Limitations:} Coefficient based approaches may not be very accurate and reliable since they rely on multiple assumptions.

\subsubsection{Similarity Analysis Based Estimation}
The principle behind this approach is to find the similar configuration(s) as the entity (VM, container, etc.) whose energy we are trying to estimate, in an existing dataset or benchmark and use its power characteristics for the energy estimation. For instance, an experiment conducted by the Teads Engineering Group followed the same core idea. They selected machines that had similar characteristics to an AWS \textit{c5.metal} instance (VM) in terms of attributes (like CPU generation, family, frequency, threads, TDP, and total memory) from the SPECpower benchmark \cite{teads,spec}. The power curve of these selected machines is then utilized to estimate VM energy. \textbf{Limitations:} Since it might not be feasible to find the exact match every time, this approach may not be very accurate in its estimation.

\subsubsection{Analytical Models Based Estimation}
 An analytical power model defines the relationship between a system's power consumption and other factors that impact the same. Hardware power models employ factors like fan speed, voltage, current, etc., to predict resultant power. They also require extensive sensors to get the input data and are difficult to scale. On the other hand, software models are widely researched and utilize metrics such as CPU utilization, memory utilization, number of cache misses, etc. Multiple literature sources \cite{10.1145/3390605}, \cite{8807458} exist that delve deeper into server power modeling to predict data center energy, while others \cite{Uchechukwu2014EnergyCI} focus on modeling cooling and power conditioning systems' energy consumption. With the rising demand for Machine Learning and Generative AI, the green community is developing software based analytical power models for these new domains \cite{faiz2023llmcarbon},  \cite{10.1145/3604930.3605705}.  \textbf{Limitations:} Most analytical models leverage linear energy curves for servers/VMs, which do not model their power characteristics well and hence are less accurate.

\subsubsection{Custom Model Training Based Estimation}
This category encompasses the generation of your own models based on certain hardware or software characteristics. One of the ways to create these power models is to generate a data set first. For instance, Fan et al. \cite{10.1145/1250662.1250665}  explain how for every machine family, they run certain representative workloads. Then the resultant system power is measured against CPU utilization, and a curve is predicted that best models the measured observations. This power curve can then be utilized to estimate power for a new set of machines. 
In another use case, an already existing benchmark can be utilized to generate a power model like in \textit{ESAVE} \cite{10.1145/3551349.3561170}. \textbf{Limitations:} This approach needs significant time and resources to develop the custom models.

\subsection{Black-Box Approaches}
This includes the approaches such as CSP based tools that do not disclose their internal calculation methodology due to proprietary and other reasons. For example, Amazon Web Services (AWS) provides \textit{Customer Carbon Footprint Tool} for its accounts \cite{awscustomercarbon}. The tool allows its end stakeholders to analyze carbon data at a multi-dimensional level with month, service, and geography. It also highlights the impact of Amazon-enabled renewable projects on the emissions. Similarly other tools include Microsoft's \textit{Emissions Impact Dashboard} \cite{emissionsimpactdashboard}, Google Cloud's \textit{Carbon Footprint} \cite{googlecloudcarbonfootprint}, \textit{Azure Machine Learning} \cite{azuremachinelearning}, \textit{AWS CloudWatch Agent} \cite{awscloudagent}.  To the best of our knowledge, we are not aware of the actual tools/approaches undertaken by the tool at the finer granular levels. \textbf{Limitations:} These tools are available to only the clients of these CSPs and do not allow major customization.

This proposed taxonomy provides a software practitioner with a concise yet thorough breadth of knowledge regarding the energy and carbon calculation approaches that exist currently. It enables her to select which category of approach can best suit their particular use case and in which direction (depth) further efforts should be put.

\section{Tools for Energy and Carbon calculation }\label{tools}

During our study, we came across multiple tools that function at different levels of granularity and with varied approaches for energy and carbon calculation. We present an exploration of some of the most relevant tools from our survey. 

\textbf{RAPL:}
Running Average Power Limit (RAPL) is designed to measure the power consumption of CPU, GPU (integrated), and RAM domains in real time \cite{khan2018rapl}. Introduced in the Intel Sandy Bridge architecture, RAPL allows energy monitoring at fine granularity and a high sampling rate \cite{khan2018rapl}. RAPL has different power domains available for Linux systems and the associated components that enable	energy consumption. For example, Package (PKG) measures the energy of the entire socket, Power Plane 0 (PP0) is for all processor cores on the socket, etc. RAPL's energy counters can be accessed through model-specific registers (MSRs). For example,  MSR\textunderscore PKG\textunderscore ENERGY\textunderscore STATUS holds the energy readings for package domain (PKG). RAPL is used by the majority of tools in our survey.

\textbf{Turbostat:}
Turbostat is provided by the Linux kernel-tools package and provides processor topology, frequency, idle power-state statistics, temperature, and power usage on Intel® x86 processors \cite{turbostatlinux}. The power information is read using Intel RAPL MSRs. Turbostat can measure CPU and RAM energy consumption on laptops, desktops, servers, and VMs (only bare metal instances).

\textbf{Intel Power Gadget:}
It is a software based utility for power usage monitoring and is enabled for Intel Core processors (from 2nd up to 10th Generation) \cite{prieto2022energy}. Intel Power Gadget supports Windows and MacOS operating systems. It uses RAPL as its underlying interface and covers CPU and RAM components for laptop and desktop machines.

\textbf{pyJoules:}
This tool measures the energy of a machine while it is executing a Python code \cite{pyJoules}. It uses Intel RAPL to monitor the power consumption of CPU, RAM and integrated Intel GPUs. For dedicated Nvidia GPUs, it leverages NVIDIA Management Library (NVML).  NVML is a C-based API for monitoring and managing different states of the NVIDIA GPU devices, exposed via nvidia-smi \cite{nvmlDeviceGetPowerUsage}. NVML enables access to  \textit{Power Management} state that provides the current board power draw. pyNVML is a Python wrapper around the NVML library \cite{pynvml}.   PyJoules calculates global energy consumption, including all background processes on the machine.

\textbf{CodeCarbon:}
It measures the carbon footprint produced by the hardware during the training and deployment of AI models (Python code) \cite{codecarbon}. The tool enables the energy consumption of different components like GPU (using pynvml), CPU (using Intel Power Gadget), and RAM (using a fixed constant of 3 Watts for 8 GB). CodeCarbon uses \textit{Carbon Intensity}, which is calculated as a weighted average of emissions from different sources like fossil fuels and green electricity (renewables). It uses carbon intensity values based on the cloud provider or country. If unavailable, it uses fixed values for energy sources or a global average of 475 gCO2.eq/KWh.

\textbf{eco2AI:}
The eco2AI library, available as a Python package, evaluates carbon emissions during the training phase of ML models \cite{budennyy2022eco2ai}. This falls under the category of both Monitoring and Estimation tools since it uses pyNVML for GPU monitoring and TDP/static coefficients for CPU and RAM components. A detailed data set of 3279 Intel and AMD processors with their TDP values enables the CPU energy estimation process, while eco2AI maintains a similar database of emission intensity coefficients for various countries/regions. 

\textbf{MLCO2 Impact:}
This is an online estimation tool that predicts the carbon impact of running a software code \cite{mlco2impact}. The tool takes user input such as hardware type (CPU/GPU type), the infrastructure information like the CSP details, the region where the hardware is deployed, and the total runtime of compute in hrs. Using TDP and Flops information, the tool outputs the resultant carbon emission and the offset values for a particular CSP. Carbon offsetting means reducing the GHG emissions by investment in renewable energy projects, restoration of trees, etc. 

\textbf{Experiment-Impact-Tracker:}
This monitoring tool gathers metrics like Python packages and version numbers, CPU/GPU hardware information, and their per-process utilization, etc., to track energy consumption, carbon emissions, and system utilization. It leverages RAPL and Intel Power Gadget for CPU and RAM while \textit{nvidia-smi} for GPU power draw. nvidia-smi is a command line utility on top of NVML. It takes special care to ensure that per-process energy is calculated by measuring per-process CPU utilization using \textit{psutil} and per-process GPU utilization using nvidia-smi \cite{henderson2020towards}.  

\textbf{CarbonTracker:}
As a software based monitoring tool, CarbonTracker calculates and predicts the energy consumption and carbon footprint of components like CPU, GPU, etc. while training deep learning models. It's forecasting feature that predicts carbon emissions using linear models, gives the end stakeholder sufficient control to stop the model training in case the environmental cost thresholds are expected to cross. It enables a plug-and-play experience by supporting clusters, desktop computers, and Google Colab notebooks  \cite{anthony2020carbontracker}.

\textbf{Cloud Carbon Footprint:}
This application aims to provide a location-based cloud emission estimate for public cloud usage for CSPs like AWS, Google Cloud, and Microsoft Azure \cite{ccf}. The application pulls this usage data from different CSPs and provides the resultant energy and carbon data at different granularity that can be filtered out by cloud provider, account, service, and time period. For CPU and storage emissions, it leverages Etsy's Cloud Jewels methodology (Coefficient based), while it uses chosen coefficients for Memory and Networking. For CPU, this uses a simple linear model (Analytical based) to estimate \textit{Average Watts}, while the average vCPU Utilization is fetched from cloud provider APIs or defaults to 50\%. The tool also takes \textit{Embodied Emissions} into account.

\textbf{Cumulator:}
This tool aims to quantify and report the carbon footprint in the context of ML models. For GPU's energy and carbon footprint estimation, it considers TDP as \textit{cost of computations}. It also considers the \textit{cost of communication} that is based on \textit{The Shift Project's 1 byte model} \cite{onebyte}.For carbon emissions, an average carbon intensity value is used \cite{trebaol2020cumulator}.  

\textbf{Tracarbon:}
It was initially built to calculate energy and carbon emissions of a device \cite{tracarbon}. In later iterations, RAPL interface (Software based) was added which allowed energy estimation for different components on Linux systems. In combination with Kubernetes, Tracarbon allows energy consumption of the containers running on the host as well.

\textbf{Green Algorithms:}
This online energy and carbon estimation tool takes the runtime of an algorithm, the details of the hardware on which it is run and the platform (cloud/personal computer etc) used for the computations, as input. It also requires the pragmatic scaling factor (PSF) as an input to take the number of time the algorithm is run into account \cite{lannelongue2021green}. The tool equates emissions to CO2 from a passenger car trip, a Paris-London flight, or the number of trees needed to sequester the carbon.

\textbf{PowerJoular and JoularJX:}
PowerJoular is a multi platform, command-line based power monitoring tool. It is written in Ada, and powered by Intel RAPL through the Linux Power Capping Framework, and nvidia-smi. PowerJoular can perform process-level energy consumption as well by using its PID. JoularJX built on top of PowerJoular can be used to measure the energy for Java methods and source code \cite{noureddine2022powerjoular}.

\textbf{Power Measurement Toolkit:}
PMT is a software library that enables power estimation in C++ and Python applications. Written in C++, PMT uses NVML for NVIDIA, rocm-smi for AMD GPUs, and RAPL or LIKWID \cite{5599200} for CPUs \cite{corda2022pmt}. 
ROCm System Management Interface is a C library for Linux that provides an interface for controlling and monitoring the AMD GPU kernel and corresponding applications \cite{rocmsmi}.

\textbf{Cloud Jewels:}
The aim of this project was to estimate general conversion factors called \textit{Cloud Jewels} that define the energy consumed by Compute, Storage and Network components in a cloud environment. Ultimately, these Cloud Jewels could be multiplied with the usage data to achieve a high level understanding of the energy impact of services run on the cloud. Cloud Jewels \cite{etsycloudjewels} estimates \textit{1.52 Wh/TBh for SSD storage}, \textit{0.89 Wh/TBh for HDD storage} and \textit{2.10 Wh per vCPUh for Server Compute} in a cloud environment. 

\textbf{SmartWatts}
This tool is an architecture agnostic and processor aware power monitoring system that can automatically adjust its CPU and DRAM power models. SmartWatts employs RAPL interface and is one of the few tools that can work for processes, virtual machines and containers. Once the CPU and DRAM component energy is calculated using RAPL, the Hardware Performance Counters (HwPC) are used to measure the container's resource usage. Based on this resource usage, the total energy consumption is distributed across the containers (Docker containers, Kubernetes pods, etc.) \cite{fieni2020smartwatts}.

\section{Analysis of Explored Tools}\label{toolcomparison}
In this section we conduct a thorough analysis of the approaches and tools that we have covered in our survey. They represent the current state-of-the-art and form the foundation for understanding the impact of green optimization strategies. 

\subsection{Comparative Study of State-of-the-Art}
We compare and contrast different tools across multiple specific attributes that have been carefully considered. We call this resultant structure the \textit{Tool Matrix}. The attributes used for comparison are detailed in Table I, and results are presented in Table II.

\begin{table}
	\caption{Attributes for Tool Comparison and their Description}
	\tiny
	\label{tab:freq2}
	\begin{tabular}{|p{2cm}|p{5.8cm}|}
		\hline
		\textbf{Attributes} & \textbf{Description} \\ \hline
		Tools &  Denotes the name of the tool to be analysed 
		\\ \hline
		License Type & How a particular software code can be used, and distributed in the purview of legal guidelines.
		\\ \hline 
		Explicit AI Context & Defines whether the tool explicitly calculates AI energy and carbon as its core functionality.
		\\ \hline
		Category & High-level category of the tool. In the table we denote the column as \textit{M/E/BB)}. Marked as \textit{M} for Monitoring, \textit{E} for Estimation, and \textit{BB} for Black-box approaches.
		\\ \hline
		Approach & Specifies sub-category of the tool as described in Figure 1. The entries are marked as \textit{HB} for Hardware Based Monitoring tools, \textit{SB} for Software Based Monitoring,  \textit{CB} for Coefficient Based Estimation,  \textit{SAB} for Similarity Analysis Based Estimation, \textit{AMB} for Analytical Model Based Estimation, \textit{CMTB} for Custom Model Training Based Estimation. 
		\\ \hline
		Energy (CPU, GPU, RAM, Storage, Network)& Specifies which approach the tool leverages for each component 
		\\ \hline
		Carbon Estimation &  Does the approach support conversion of energy to carbon equivalent. If Yes, we specify the approach used for carbon estimation (real-time grid intensities, static long-term averages, etc.
		\\ \hline
		Working Mode & Denotes how tool works (by embedding inside python code, as a web-based calculator, etc.)
		\\ \hline
		Supported OS & Specifies if the tool is compatible with Windows, Linux, and Mac OS.
		\\ \hline
	\end{tabular}
\end{table}

\begin{table*}
	\centering
	\caption{Comparison of Explored Tools (Tool Matrix)}
	\tiny
	\begin{tabular}{|p{1.25cm}|p{0.9cm}|p{0.7cm}|p{0.6cm}|p{0.6cm}|p{1.5cm}|p{1.5cm}|p{1.5cm}|p{0.5cm}|p{0.5cm}|p{1.5cm}|p{1.4cm}|p{1.6cm}|}
		\hline
		\multirow{2}{*}{\centering \textbf{Tools}} &
		\multirow{2}{*}{\textbf{License Type}} &
		\multirow{2}{*}{\textbf{AI Context}} &
		\multirow{2}{*}{\centering \textbf{Category}} &
		\multirow{2}{*}{\centering \textbf{Approach}} &
		\multicolumn{5}{c|}{\textbf{Energy}} &
		\multirow{2}{*}{\centering \textbf{Carbon Estimation}} &
		\multirow{2}{*}{\centering \textbf{Working Mode}} &
		\multirow{2}{*}{\centering \textbf{Supported OS}} \\  \cline{6-10}

		&  &   & & & \centering\textbf{CPU} &  \centering\textbf{GPU} & \centering\textbf{RAM} & \centering\textbf{Storage} & \centering\textbf{Network} & & & \\
		\hline
		pyJoules &  MIT &  No &  M & SB &  Intel RAPL &  pynvml,
		Intel RAPL (Integrated GPU)  &  Intel RAPL &  No &  No & No & Embed inside Code & Linux \\ 
		\hline
		CodeCarbon & MIT  & Yes & M,E & SB,CB & Intel Power Gadget, RAPL, TDP-Based(Fallback) & pynvml & Static Cofficients & No & No & Static Grid Intensities, CO2 Signal & Embed inside Code & Windows, Linux, MacOS 
		\\ \hline
		eco2AI & Apache 2.0  & Yes & M,E & SB,CB & TDP-Based, Static Coefficients & pynvml & Static Coefficients & No & No & Static Grid Intensities & Embed Inside Code  & Windows, Linux, MacOS \\ \hline
		ML CO2 Impact & MIT  & Yes & E & CB & TDP Based  & TDP + FLOPS Based  & No & No & No & Static Grid Intensities & Web-Based Tool  & Web-Based Tool \\ 
		\hline
		Green Algorithms & CC BY 4.0  & No & E & CB &TDP & TDP & Static Coefficient & No & No & Data from CarbonFootprint \cite{gacarbonfootprint} & Web-Based Tool, Editable Codebase (for HPC cluster) & Web-Based Tool
		\\ \hline	
		Cumulator & MIT  & Yes & E & CB &- & TDP (Static)  & No & No & Static & Static Grid Intensities-Europe:447 gCO2eq/ kWh & Embed Inside Code, WebApp & Windows, Linux, MacOS 
		\\ \hline
		Cloud Carbon Footprint & Apache 2.0 & No & E & AMB,CB & Min/Max Watts$\vert$vCPU Hours, Average Watts$\vert$vCPU Hours (fallback) & Min/Max Watts$\vert$GPU Hours,Average Watts $\vert$ GPU Hours (fallback) & Static & Static & Static & Static Grid Intensities & Standalone app  & Windows, Linux, MacOS \\ \hline
		Power Measurement Toolkit  & Apache 2.0 & No & M & SB & RAPL, LIKWID & NVML, Rocm-smi (AMD) & No & No & No & No & Embed Inside Code  & Linux \\ \hline
		Carbon Tracker & MIT  & Yes & M & SB & Intel RAPL & NVML & Intel RAPL & No & No & Energi Data Service (Denmark), Carbon Intensity API (Great Britain), Avg defaults for other regions & Embed Inside Code
		& - \\ \hline
		Scaphandre  & Apache-2.0  & No & M & SB & Powercap-RAPL (GNU/ LINUX), MsrRAPL (Windows) & No & Powercap RAPL (GNU/LINUX), MsrRAPL (Windows) & No & No & No & Terminal & Windows, Linux \\ \hline
		Experiment-Impact-Tracker & MIT  & Yes & M & SB & Intel RAPL, Intel Power Gadget & nvidia-smi  & Intel RAPL, Intel Power Gadget & No & No & Open-source data from Electricity Maps | Static Data | Realtime Data for California & Embed Inside Code  & Linux, MacOS (only CPU and memory-related metrics) 
		\\ \hline
		Tracarbon & Apache 2.0 & No & M & SB & Intel RAPL & Intel RAPL & Intel RAPL & No & No & Real time using C02Signal \cite{co2signal}/ Electricity Maps \cite{electricitymaps},	Static Grid Intensities for Europe, Static data for AWS Grid emissions factors from CCF	& Embed Inside Code, CLI  & Linux (Only RAPL based), MacOS (Global energy, only if plugged into wall adapter )
		\\ \hline
		SMART-WATTS  & BSD-3-Clause & No & M & SB &  RAPL$\vert$HwPC (processes, VMs, containers)  & No & RAPL$\vert$HwPC (processes, VMs, containers)  & No & No & No & power meter is implemented as a software service that can be deployed on a single node.  & Linux
		\\ \hline
		PowerJoular and JoularJX & GPL-3.0  & No & M & SB & RAPL & nvidia-smi
		& RAPL & No & No & No & command
		line, System service & Linux, JoularJX uses a custom program monitor (based on Intel Power Gadget) to run on Windows.
		\\ \hline
		Azure Machine Learning  & Azure tool  & Yes & BB & - & -  & GpuEnergyJoules metric provides the relevant data at intervals of one min.  & No & No & No & No & Web-Based Tool /Managed End points  & Web-Based Tool
		\\ \hline
		Amazon CloudWatch  & AWS tool  & Yes & M & SB & No & Integrate your DLAMI with the unified CloudWatch agent or by using preinstalled gpumon.py script. power\_draw is the relevant metric.
		& No & No & No & No & Command line/system
		service  & Linux  \\ \hline
		Cloud Jewels  & MIT   & No & E & CB & Sources include  U.S. Data Center Energy Usage Report, The Data Center as a Computer, and the SPEC power report.& -& No & Same as CPU & No & No & Fixed Coefficients  & Windows, Linux , MacOS 
		\\ \hline
		DNN Energy Estimation Tool  & -  & Yes & E & - & \multicolumn{5}{p{7cm}|}{Multiply-and-accumulate (MAC)s and memory access energy-based approach.} & No & Web-Based Tool  & Web-Based Tool  \\ \hline
		Carbon Footprint  & GCP tool & No & BB & - & \multicolumn{5}{|p{7cm}|}{Reports are prepared according to the Greenhouse Gas Protocol (GHGP) carbon reporting (location-based, market-based), accounting standards (GHGP) and apportions its emissions into Scope 1,2 and 3 which are further accounted as part of GCP customer's Scope 3 emissions. Google separately evaluates the energy used when running a workload ("dynamic power") against the energy used when machines are idle ("idle power") and builds its calculations from bottom-to-top. It then uses hourly greenhouse gas emission factors to calculate location-based emissions (electricity maps), while annual greenhouse gas emission factors are used to calculate market-based emissions}
		& Google’s scope 1, 2, and 3 & Web-Based Tool   & Web-Based Tool   \\ \hline
		Emissions Impact Dashboard  & Microsoft tool  & No & BB & - & \multicolumn{5}{|p{7cm}|}{Follows Greenhouse Gas Protocol and emissions associated with Scope 1,2 and 3. Power consumption for equipment at different utilization was validated using SPECpower\_ssj\_2008 dataset. Scope 2 calculation takes into account DC/server efficiency, grid emission factors, RE purchases and power usage. Scope 3 emissions are calculated at individual component and IT hardware level and finally customer emissions are connected to per service per region.}
		& Microsoft’s scope 1, 2, and 3 & Web-Based Tool   & Web-Based Tool
		\\ \hline   
		Customer Carbon Footprint Tool  & AWS tool    & No & BB & - &  \multicolumn{5}{|p{7cm}|}{The tool adheres to the Greenhouse Gas Protocol and ISO 14064, and include Scope 1 and Scope 2 emissions. Scope 2 portion of the estimate is calculated using the GHGP market-based method.}& Scope 1 and Scope 2 emissions  & Web-Based Tool (AWS Console)  & Web based tool.  \\ \hline
	\end{tabular}
	\label{tab:sample}
\end{table*}

This representation (Tool Matrix) empowers one to evaluate which current tool or set of tools is applicable for a particular use-case based on available telemetry and level of intrusiveness. The end stakeholders in this space can also identify the \textit{top tools} in the domain that are applicable across different scenarios. Finally, for the practitioners, who are aiming to build their own tool, this matrix helps them understand the different nuances and capabilities that should be part of an \textit{``ideal"} energy and carbon calculation tool. 

\subsection{Component wise Consolidation of Energy and Carbon Calculation Approaches}
Often, we need to focus on the energy calculations for a particular component like CPU, GPU, RAM, Storage and Network, making it necessary to have a compilation of different approaches that exist in the literature. In Table III, we present this consolidation of approaches along with a few examples of tools that implement the same.

\begin{table}
	\caption{Component wise Energy and Carbon Calculation Approaches}
	\tiny
	\label{tab:freq}
	\begin{tabular}{|p{1cm}|p{1cm}|p{2.25cm}|p{3cm}|}
		\hline
		\centering \textbf{Measured Components} & \centering \textbf{Monitoring/ Estimation}& \centering \textbf{Approaches} &  \textbf{Example of Tools}
		\\ \hline
		\multirow{5}{*}{CPU} & Monitoring & Software Based &  CodeCarbon, Experiment-impact-tracker 
		\\ \cline{2-4}
		& \multirow{4}{*}{Estimation} & Coefficient Based & eco2AI, MLCO2Impact 
		\\ \cline{3-4}
		& & Similarity Analysis Based & Teads Approach
		\\ \cline{3-4}
		& &Analytical Models Based & Cloud Carbon Footprint 
		\\ \cline{3-4}
		& &Custom Model Training Based & ESAVE 
		\\ \hline
		\multirow{3}{*}{GPU} & Monitoring & Software Based &  pyJoules, PMT
		\\ \cline{2-4}
		& \multirow{2}{*}{Estimation}& Coefficient Based & Cumulator ,Green Algorithms		
		\\  \cline{3-4}
		& & Analytical Models Based & Cloud Carbon Footprint
		\\ \hline
		\multirow{2}{*}{RAM} & Monitoring & Software Based &  CarbonTracker, PowerJoular
		\\ \cline{2-4}
		& Estimation & Coefficient Based & eco2AI, Green Algorithms
		\\ \hline
		Storage & Estimation &Coefficient Based & Cloud Jewels
		\\ \hline
		Network & Estimation &Coefficient Based & Cumulator	
		\\ \hline
	\end{tabular}
\end{table}

\section{Challenges}\label{challengesshort}
Although the current tools allow the end stakeholders to calculate their software's energy consumption, they still suffer from a few drawbacks. Some of these are elaborated below:-

\textbf{Lack of visibility into process level energy consumption:}
	 Most tools estimate or monitor energy consumption at the system level that encompasses backend operations and concurrently running processes on the same machine. Process-level energy consumption, which is crucial for optimizing energy consumption at granular levels, is not provided by most of the existing tools. For instance, CodeCarbon, when configured with \textit{tracking\_mode = Process}, currently estimates the energy for RAM at the process level and not the GPU/CPU energy. 
	 
\textbf{Limited use of real time values for RCI:}
	 Many tools employ static long-term averages for RCI. However, RCI may vary by more than 70\% in a day depending on the grid's energy source (renewable/non-renewable) \cite{rcidata}. For example, on a particular day in the US Ohio location (which correlates to the AWS region us-east-ohio), the RCI value varied from $41.5 ktCO_2eq/kWh$ to $73.9 kt_CO2eq/kWh$ in 24 hours. Similarly, RCI values vary significantly across different days, months, and the time of day a model is trained. Thus, carbon calculation tools must use real-time grid intensities for high accuracy.
	 
\textbf{Neglecting Embodied Emissions:}
	  When we run a particular application on a device, this execution must include a fraction of the device's total embodied emissions \cite{ccfembodied}. Most tools do not factor in embodied emissions while estimating the total carbon emissions. Recently embodied emissions have dominated computing's carbon footprint \cite{10.1145/3470496.3527408}. Hence, when we miss out on this important component in carbon estimation, it has a direct impact on subsequent sustainability-led choices. 
	  
\textbf{Dependence on approximate approaches:}
	  For estimating CPU energy consumption, the majority of tools still rely upon the TDP value of the device. TDP is not the maximum power the processor can consume and is not a good estimate of energy \cite{inteltdppaper}. Moreover, many tools do not take the actual CPU utilization during a particular task into account while estimating the energy consumption.  50\% of the TDP is considered to be the average power consumption. This can significantly impact the estimation accuracy. A better way is to observe the CPU utilization during the entire process and multiply it with TDP and time to estimate the energy numbers.
	  
 \textbf{Lack of reliable estimation approaches for GPU:}
	   Current tools use the pyNVML library to monitor the GPU energy consumption, which requires access to the underlying machine. But estimation based tools are not only less intrusive but can also estimate the energy consumption before or without the execution of software. Currently, only TDP based estimation approaches are available, which may not be very accurate in all scenarios.
	  
\textbf{Lack of GPU agnostic energy monitoring tool:}
	   The vast majority of energy monitoring tools do not work for all types of GPUs. While many leverage NVML for NVIDIA GPUs, few employ ROCm System Management Interface for AMD ones. There is certainly an absence of a single tool that can support all types of GPU hardware and make the process of energy calculation more seamless and platform-agnostic.
	  
\textbf{No reporting of SCI:}
	   The SCI value endorsed by the Green Software Foundation helps measure the carbon impacts of software systems. SCI has recently achieved the ISO standard and has become an important metric that takes energy efficiency and carbon awareness into account \cite{sciiso}. For example, while comparing multiple LLM models in performing certain tasks (with similar prompts), SCI scores can be a good indicator of their environmental impact. Currently, most tools do not report SCI, while a few, like Cloud Carbon Footprint, aim to adhere to the standard effectively \cite{ccfembodied}.
	   
\textbf{Dearth of reliable approaches for components other than Compute:}
Our survey revealed that mainly coefficient based approaches are available for energy calculation of components like Network, Storage, and Memory. While this may work for a few scenarios, the energy contribution of these components becomes predominant in certain cases. For instance, EC2 High Memory (U-1) instances are specially designed to process large datasets in memory. Hence, in such cases, we need more reliable and accurate approaches for these components.
	   
\textbf{Lack of Multi-Experiment Analysis Dashboard :}
	   Experiment tracking is crucial in software development, especially in the context of training ML models. As the projects and team size scale, tracking experiment metadata (e.g., model types, hyper parameters) becomes challenging, especially across multiple iterations by different team members. Most existing tools lack multi-experiment tracking and visualization capabilities. CodeCarbon provides this feature by integrating with Comet \cite{codecarboncomet}. In the future, it will be useful to have these features already built into energy calculation tools.

Our motivation behind bringing forth these limitations is to guide the green software community towards new research areas that need attention and effort.
\section{Conclusion and Future Directions}\label{conclusionfuturedirections}

Understanding and optimizing the environmental impact of software needs energy and carbon calculation approaches/tools as the first step. This paper presents a state-of-the-art study that compares and contrasts 21 tools across different attributes. We provided a component wise (CPU, GPU, RAM, Storage, Network) consolidation of energy and carbon calculation approaches. Our proposed taxonomy (with identified drawbacks) and further categorization along with the \textit{Tool Matrix} help stakeholders evaluate suitable tools for specific use cases. We highlight key challenges in the current space, which serve as important pillars for further research in this domain. In the future, we plan to assess these tools' \textit{accuracy} through controlled experiments and expand the study to include more tools. Based on our \textit{Tool Matrix}, we are developing an initiative that intelligently recommends the most suitable energy and carbon calculation tool for a given user scenario. To mitigate a few of the existing challenges, we are also building our own energy and carbon calculation approach for the AI life cycle. We plan to share this approach as it matures. This is a vast domain, and we encourage the green software community to collaborate in this relevant research area.

\bibliographystyle{IEEEtran}
\bibliography{IEEEabrv, Bibliography} 

\end{document}